\documentclass[aps,nofootinbib,showpacs]{revtex4}
\usepackage{amsmath}
\usepackage{amsfonts}
\usepackage{amssymb}
\usepackage{graphicx}
\usepackage{epsf}
\newcommand{\ud}{\mathrm{d}}

\begin{document}
\title{The surface density of holographic entropy}
\author{V.V.Kiselev, S.A.Timofeev}
\affiliation{Russian State Research Center ``Institute for High
Energy Physics'',
Pobeda 1, Protvino, Moscow Region, 142281, Russia\\
Moscow Institute of Physics and Technology, Institutskii per. 9,
Dolgoprudnyi, Moscow Region, 141701, Russia}
\begin{abstract}
On the basis of postulates for the holographic description of
gravity and the introduction of entropic force, for static sources
we derive the universal law: the entropy of a holographic screen
is equal to quarter of its area in the Planck system of units.
\end{abstract}

\maketitle
\section{Introduction}

Following the thermodynamic interpretation of black hole horizon
in general relativity
\cite{Bekenstein,Bekenstein_1,Bekenstein_2,Hawking,Hawking_1,Hawking2,
Hawking_2,Hawking_3,Jacobson,Padmanabhan_1,Padmanabhan} and the
holographic principle \cite{Holograph,Holograph_1,B-rev} 
Verlinde \cite{Verlinde} and Padmanabhan \cite{Padmanabhan2}
(both independently) have shown that Newton's gravity appears by
introducing several postulates about holographic
screens\footnote{Developments of these arguments can be found in
\cite{develope,develope_1}.}. These holographic screens generate
both the space and the entropic force, which is equivalent to the
gravitational acceleration. The entropic force is treated as the
gradient of entropy, that emerges, when the probe particle is
approaching  the screen. In accordance with the holographic
principle, a theory in the three-dimensional space can be
described in terms of theory on the two-dimensional surface
covering the three-dimensional space. Thus, one dimension of the
space is ``holographically emerged'' and the information about
particles inside the surface is encoded on the surface.
Three-dimensional space is considered as a union of the
holographic screens, which are characterized by temperature and
entropy. The microscopic modes hidden on the screen are
chaotically distributed in the way that one can tell about
thermodynamic system ``bits on the surface''. In the Newton's
mechanics with the simplest case of  point source, the screen is
an equipotential surface, and in the general case the screen
corresponds to a surface of constant acceleration for free falling
bodies. Such a surface is ascribed to the constant Unruh
temperature \cite{Unruh}.

The following postulates are introduced in \cite{Verlinde}:
\begin{enumerate}
\item The displacement $\delta x$ of the particle with mass $m$ in
the emergent holographic direction perpendicularly to the screen
in its vicinity changes the screen entropy $S$\footnote{Here and throughout the
paper, we assume that $k_B$ is the Boltzmann constant, $G$
is the Newton gravitational constant, $\hbar$ is the Planck constant, $c$
is the speed of light.} according to the law
\begin{equation}
      \delta S = 2\pi k_B \frac{mc}{\hbar}\; \delta x.
\end{equation}
\item The number of the degrees of freedom (bits) are proportional
to the screen area $A$ and equal to the number of area cells in the
Planck system of units
\begin{equation}
 N=\frac{A  c^3}{G \hbar}.
\end{equation}
\item The gravitational energy is divided evenly over these
degrees of freedom,  i.e. the equipartition distribution of bits takes place.
\end{enumerate}
As shown the gravitational force arises due to the entropy
gradient, i.e. the gravity is the entropic force\footnote{The
ideas offered by Verlinde in \cite{Verlinde} have generated wide
discussions and investigations, wherefrom we can mention
applications to cosmology
\cite{iCosmo,iCosmo1,iCosmo2,iCosmo3,iCosmo4,iCosmo5,iCosmo6,iCosmo7,iCosmo8,iCosmo9,
iCosmo10,iCosmo11,iCosmo12}, entropic corrections to the Newton's
law of gravity
\cite{iNewton,iNewton1,iNewton2,iNewton3,iNewton4,iNewton5},
interpretations in terms of Rindler horizon
\cite{iRindler,iRindler1,iRindler2,iRindler3}, study of specific
gravitational systems \cite{specific,specific1,specific2},
relations with Loop Quantum Gravity \cite{iLQG,iLQG1,iLQG2} and a
specific microscopic justification of bits on the surface
\cite{Maekela}.},
\begin{equation}
      F\delta x= T\delta S.
\end{equation}

Similar ideas were independently developed by Padmanabhan \cite{Padmanabhan2}.
So, in \cite{otherPadmanabhan} he has introduced the definition of entropy
$S=\frac1{8 \pi G} \int \sqrt{-g} \ud ^4 x \nabla_i a^i$ for a static spacetime
in terms of proper acceleration $a^i$ for a comoving observer (here $\nabla_i$
is a covariant derivative). Such definition is based on the statement
that the entropy of black hole  is proportional to the area of horizon with an
appropriate coefficient especially fixed. In \cite{Padmanabhan2} Padmanabhan
has introduced the equipartition law for the horizon degrees of freedom\footnote{
Note that logical arguments and postulates used by Verlinde and Padmanabhan
are clearly different (Verlinde's statements are more simple and, hence, more
speculative, while Padmanabhan's formulation is more technical and specific),
but they lead to identical relations between relevant physical quantities.}.

On the other hand,  variants for the internal structure of black holes as
suggested by the string theory results in the unique relation between the
horizon area of a black hole and its entropy $S =k_B\ln \Omega$, calculated
over the number of equipartitional microscopic states $\Omega$ with identical
quantum numbers of the black hole for exterior observer (some charges and
the mass)
\cite{Strominger-Vafa}, namely,
\begin{equation}
 S=\frac14 \frac{k_B c^3}{G \hbar} A ,
\end{equation}
where $A$ is the horizon area. Note that the dimensionless
coefficient $\frac14$ is universal but not trivial result of a
calculation according to Cardy formula \cite{Cardy,Cardy1} in the
conformal field theory. This formula is asymptotically counting
for the amount of states with given invariants in Virasoro algebra
(a central charge and a quantum number of generator for null index
modes). It is surprising, since \textit{a priori} one could not
expect any preferable value for this factor! Similarly, in papers,
wherein solutions describing black holes are matched with
conformal symmetry algebras, the same Cardy formula gives this
factor for the entropy proportional to the area (see review in
\cite{Carlip,Carlip1,Carlip2,Carlip3}). Note also that the
thermodynamic interpretation of black hole ``equation of state'',
i.e. a relation between the black hole mass and its horizon area,
angular momentum and charges gives the same coefficient if the
horizon is assigned to the Hawking temperature. In this respect,
the natural question does arise: why do microscopically different
theories give the same simple dependence of entropy on the horizon
area? \cite{Strominger}.

It is interesting that in the Verlinde's article
\cite{Verlinde} the event horizon can be considered as the
holographic screen. Therefore, the relation between the horizon
area and its entropy gives rise to the relation between the screen
area and its entropy. Thus, we can more universally consider the
question about the entropy density on the holographic screen for
static systems using the postulates for the holographic screens
and entropic force.

Here we establish the area law for the holographic
entropy
\begin{equation}
 \frac{\ud S}{\ud A}=\frac14 \frac{k_B c^3}{G \hbar},
\end{equation}
starting from  universal principles describing the gravitational force as the
entropic force as formulated by Verlinde \cite{Verlinde}.

\section{The entropy density}

Let us consider nonrelativistic particle with kinetic  and
gravitational potential energies denoted by $K$ and $U$,
respectively. The total energy $\tilde E$ changes due to
velocity and  potential variations (due to both a particle
displacement and a potential change because of a source
variation)
$$
\ud \tilde E = \ud K + \ud U.
$$
The change $\ud U$ is associated with the gravitational forces, which
arise because of holographic screens possessed entropy. Then, it
is caused by a transfer of energy from the holographic screen to
the particle. Therefore, the holographic screen energy $E$ changes
due to the screen entropy variation and exchange by the energy, i.e.
the work of gravitational force, so that
\begin{equation}\label{diff}
      \ud E = T \ud S - \ud U.
\end{equation}
According to the Hamilton equations for the motion of a particle
$$
\delta U =\frac{\partial U}{\partial x}\delta x= - F \delta x.
$$
The introduction of the entropic force acting on the particle in
accordance with the postulate for the change of holographic screen entropy
under the particle displacement\footnote{The position of the
particle is an external parameter for the holographic
screen as the thermodynamical system. The entropy depends on this parameter
according to the postulate, so that in the thermodynamical equilibrium the
entropy extreme can be ensured by introduction of the work of the
entropic force. Indeed, the entropy as a function of energy $E$ and the
external parameter being the position of particle $x$ can be written
in the form $S=S(E+U,x)$, and the entropy is extremal,
when $\frac{\partial S}{\partial x}=0$, i.e. there is the thermodynamic
equilibrium, and $U=0$. However, the postulate about the dependence of
entropy on the particle position $\frac{\partial S}{\partial
x}\neq 0$ agrees with the thermodynamic equilibrium if  only the
entropy is extremal, hence,
$$
      \frac{\ud S}{\ud x}=\frac{\partial S}{\partial x}+\frac{\partial
S}{\partial E}\,\frac{\partial U}{\partial x}=0.
$$
Therefore, from the formulas $\frac{\partial U}{\partial x}=-F$ and
$\frac{\partial S}{\partial E}=\frac{1}{T}$ we find the relation for the
entropic force $F\delta x=T\delta S$. If we pass from the system of the
holographic screen and the particle to the open thermodynamic
system of the holographic screen, which can get an external transferred energy
due to the work related to the variation of the screen area, for example,
then for the total entropy differential $S=S(E+U,x)$ we obtain
$$
      \ud S =\frac{\partial S}{\partial E}\,(\ud E+\ud U)\quad
      \Rightarrow\quad
      \ud E=T\ud S-\ud U.
$$
That is equivalent to the derivation of (\ref{diff}) given above.}
means that
\begin{equation}\label{diff2}
      F \delta x = T \delta S,\quad
      \Rightarrow\quad
      \delta U  = -T \delta S,
\end{equation}
where $\delta S$ is the variation of holographic screen
entropy\footnote{Emphasize the following feature of gravity: When a particle
is freely falling down to the horizon, the gravitational forces make a positive
work, while the gravitational mass of system increases from the viewpoint of an
observer outside the horizon, i.e. $\delta E>0$ at  $ F\delta x=-\delta
U>0$.}.

In the thermodynamic equilibrium there is the surface  density of entropy
$\frac{\ud S}{\ud A}$, which can depend on the screen temperature, hence, the
entropy change is associated with the area variation
\begin{equation}\label{delta}
      \delta S=\frac{\ud S}{\ud A}\,\delta A,
\end{equation}
so that
\begin{equation}\label{delta U}
      \delta U=-T \,\frac{\ud S}{\ud A}\,\delta A.
\end{equation}
Thus,
\begin{equation}\label{dE}
      \ud E=T\ud S + T \,\frac{\ud S}{\ud A}\,\ud A.
\end{equation}
In this formula the differential of entropy corresponds to an
arbitrary variation, while the differential of area is  strictly given by a
small element of the surface area on the specific holographic screen. In
addition, we can ascribe the second term in (\ref{dE}) to the work of surface
tension. In this respect, one can talk on the surface tension origin of
entropic force in the case of gravity.

Then, according to (\ref{dE}) we find the surface density of entropy related
with the surface density of energy by
\begin{equation}\label{extra}
      \frac{\ud E}{\ud A}= 2 T \frac{\ud S}{\ud A}.
\end{equation}
We emphasize that physical meaning of (\ref{extra}) is clear,
namely, the energy density is determined by two terms: the contribution of the
entropy density and the work performed, when the surface of holographic screen
is changed\footnote{This conclusion is analogous to the determination of
energy density $\epsilon$ for a gas, when the relation $\ud E = T \ud S - p\,
\ud V$ straightforwardly yields
$\epsilon=\frac{\ud E}{\ud V}=T \frac{\ud S}{\ud V} -p$, where $p$ is the
pressure.}, whereas these contributions are equal to each other. In other words,
considering the part of the holographic screen with a larger area corresponds
to an increase of the energy of thermodynamic subsystem,  and it is accompanied
by both the increase of the subsystem entropy and making the positive work in
order to increase the subsystem area.

Thus, the equipartition distribution of modes on the holographic screen, i.e.
$E=\frac12 k_B TN$, in the differential form gives
$$
\frac{\ud E}{\ud A} =\frac{1}{2}\, k_BT \frac{c^3}{G\hbar},
$$
and it leads to the constant value of  the surface density of entropy
\begin{equation}\label{main}
      \frac{\ud S}{\ud A} = \frac14 \frac{k_B c^3}{G \hbar}.
\end{equation}
Note that in the  derivation above we have not used the explicit
definition of temperature for the holographic screen and the postulate
on the  entropy change under the particle
displacement near the screen. These quantities become significant
only for the calculation of the explicit dependence of gravitational forces on
the distance.

The calculated constant surface density of entropy for the holographic
screen allows us to determine the equality of the black
hole entropy to a quarter of its horizon area in the Planck system of
units. Note that there is the difference between a holographic screen
and a black hole horizon. The holographic screen of
gravitating system with a given mass can be arbitrarily posed and
its area is variable independent of the system mass, while the
black hole horizon is determined by its mass and other charges, and
the horizon area can be evaluated according to ``the state equation'', i.e. the
relation of the black hole mass to its charges and the horizon
area. ``The state equation'' determines another approach to the
variation of area. At the equipartition distribution of bits on the holographic
screen we can select a part of the screen
and further relate it with an appropriate part of the gravitating energy.
In contrast, the variation of the black hole horizon area has another meaning,
since it determines the variation of the gravitating mass. However, such
difference does not affect the validity of law for the surface density of
entropy (\ref{main}). In other words,  in the case
of black hole horizon, the area variation at the fixed
temperature in formula (\ref{dE})  should be set to zero $\ud A=0$, and one
should substitute the actual relation $S=\frac14 \frac{k_B c^3}{G \hbar} A_H$,
where $A_H$ is the horizon area.

\section{Conclusion}

Thus, in the framework of the equipartition
postulate and the entropic interpretation of gravitational
force we have deduced the universal law for the surface
density of entropy for the holographic screen without setting any
microscopic structure of the screen. It means that such a
consideration is completely thermodynamic. The holographic screen  can 
particularly be associated with the black hole horizon, then it becomes the event
horizon, and one can observe its thermodynamic properties caused by the
microscopic structure of the black hole in the form of Hawking
radiation, for example. However, the holographic nature of the horizon leads to
the simple relation of the entropy to the horizon area independently of the
microstructure of the theory.

It is worth noting that formula (\ref{main}) holds in the static system
and for the static holographic screens. So, a transformation to another
inertial moving coordinate system without acceleration and inertia forces
leads to that the screen should be deformed due to the Lorentz
contraction and its area is kinematically changed. However, the screen entropy
cannot be
changed by this transformation because the number of bits on the screen cannot be 
changed in this process (see also the paper by Morozov
\cite{Morozov-kin}). Therefore, it is necessary to introduce a correction in the
second order of velocity to the constant $\frac14 \frac{k_B c^3}{G \hbar}$ in
the moving frame.

In the way of derivation offered here, we do not
see any obstacle to generalize the considered nonrelativistic
consideration up to the case of general relativity for  static
gravitational fields in the way realized in \cite{Verlinde}.

Finally, we note that relation (\ref{main}) was independently
obtained by Padmanabhan in \cite{otherPadmanabhan}, by using the
special definition of entropy in terms of the gravitational
acceleration and temperature as mentioned in the Introduction. Then,
he derived the expression for the total energy in the form
\mbox{$E_H=2T_HS_H$} for the black holes. Combining this relation
with the equipartition distribution
\cite{Padmanabhan,otherPadmanabhan,Padmanabhan2} $E=\frac12 T A_H
\frac{k_B c^3}{G \hbar}$, one can elementary find (\ref{main}) for
the fixed equation of state. In a different way, the relation
between the entropy and area was found in \cite{other} for the
event horizon, only.

This work was partially supported by grants of Russian Foundations for Basic
Research 09-01-12123 and 10-02-00061, Special Federal Program ``Scientific and
academics personnel'' grant for the Scientific and Educational Center
2009-1.1-125-055-008, ant the work of T.S.A. was supported by the
Russian President grant MK-406.2010.2.


\begin{thebibliography}{}
\bibitem{Bekenstein}
J.~D.~Bekenstein,
  ``Black Holes And The Second Law,''
  Lett.\ Nuovo Cim.\  {\bf 4}, 737 (1972);
\bibitem{Bekenstein_1}
 J.~D.~Bekenstein,
  ``Black Holes And Entropy,''
  Phys.\ Rev.\ D {\bf 7}, 2333 (1973);
\bibitem{Bekenstein_2}
 J.~D.~Bekenstein,
  ``Generalized Second Law Of Thermodynamics In Black Hole Physics,''
  Phys.\ Rev.\ D {\bf 9}, 3292 (1974).
\bibitem{Hawking}
S.~W.~Hawking,
  ``Particle Creation By Black Holes,''
  Commun.\ Math.\ Phys.\  {\bf 43}, 199 (1975);
\bibitem{Hawking_1}
 J.~B.~Hartle and S.~W.~Hawking,
  ``Path Integral Derivation Of Black Hole Radiance,''
  Phys.\ Rev.\ D {\bf 13}, 2188 (1976).
\bibitem{Hawking2}
S.~W.~Hawking,
  ``Black Holes And Thermodynamics,''
  Phys.\ Rev.\ D {\bf 13}, 191 (1976);
\bibitem{Hawking_2}
 G.~W.~Gibbons and S.~W.~Hawking,
  ``Cosmological Event Horizons, Thermodynamics, And Particle Creation,''
  Phys.\ Rev.\ D {\bf 15}, 2738 (1977);
\bibitem{Hawking_3}
 S.~W.~Hawking and G.~T.~Horowitz,
  ``The Gravitational Hamiltonian, action, entropy and surface terms,''
  Class.\ Quant.\ Grav.\  {\bf 13}, 1487 (1996)
  [arXiv:gr-qc/9501014].
\bibitem{Jacobson}
  T.~Jacobson,
  ``Thermodynamics of space-time: The Einstein equation of state,''
  Phys.\ Rev.\ Lett.\  {\bf 75}, 1260 (1995)
  [arXiv:gr-qc/9504004].
\bibitem{Padmanabhan_1}
  T.~Padmanabhan,
  ``Thermodynamical Aspects of Gravity: New insights,''
  Rept.\ Prog.\ Phys.\  {\bf 73}, 046901 (2010)
  [arXiv:0911.5004 [gr-qc]];
\bibitem{Padmanabhan}
 T.~Padmanabhan, ``Surface Density of Spacetime Degrees of Freedom
 from Equipartition Law in theories of Gravity,''
 [arXiv:1003.5665[gr-qc]].
\bibitem{Holograph}
 G.~'t Hooft,
  ``Dimensional reduction in quantum gravity,''
  [arXiv:gr-qc/9310026];
\bibitem{Holograph_1}
 L.~Susskind,
  ``The World as a hologram,''
  J.\ Math.\ Phys.\  {\bf 36}, 6377 (1995)
  [arXiv:hep-th/9409089].
\bibitem{B-rev}
 R.~Bousso,
  ``The holographic principle,''
  Rev.\ Mod.\ Phys.\  {\bf 74}, 825 (2002)
  [arXiv:hep-th/0203101].
\bibitem{Verlinde}
  E.~P.~Verlinde,
  ``On the Origin of Gravity and the Laws of Newton,''
  arXiv:1001.0785 [hep-th].
\bibitem{Padmanabhan2}
  T.~Padmanabhan,
  ``Equipartition of energy in the horizon degrees of freedom and the emergence
  of gravity,''
  arXiv:0912.3165 [gr-qc].
\bibitem{develope}
  R.~Banerjee and B.~R.~Majhi,
  ``Statistical Origin of Gravity,''
  arXiv:1003.2312 [gr-qc];
\bibitem{develope_1} Yu-Xiao Liu, Yong-Qiang Wang and Shao-Wen Wei,
``Temperature and Energy of 4-dimensional Black Holes from
Entropic Force'', arXiv:1002.1062[hep-th].
\bibitem{Unruh}
  W.~G.~Unruh,
  ``Notes on black hole evaporation,''
  Phys.\ Rev.\  D {\bf 14}, 870 (1976).
\bibitem{iCosmo}
R.~G.~Cai, L.~M.~Cao and N.~Ohta,
  ``Friedmann Equations from Entropic Force,''
  Phys.\ Rev.\  D {\bf 81}, 061501 (2010)
  [arXiv:1001.3470 [hep-th]];
\bibitem{iCosmo1}
 F.~W.~Shu and Y.~Gong,
  ``Equipartition of energy and the first law of thermodynamics at the apparent
  horizon,''
  arXiv:1001.3237 [gr-qc];
\bibitem{iCosmo2}
 M.~Li and Y.~Wang,
  ``Quantum UV/IR Relations and Holographic Dark Energy from Entropic Force,''
  Phys.\ Lett.\  B {\bf 687}, 243 (2010)
  [arXiv:1001.4466 [hep-th]];
\bibitem{iCosmo3} Y.~Wang,
 ``Towards a Holographic Description of Inflation and Generation of
 Fluctuations from Thermodynamics,''
  arXiv:1001.4786 [hep-th];
\bibitem{iCosmo4} D.~A.~Easson, P.~H.~Frampton and G.~F.~Smoot,
  ``Entropic Accelerating Universe,''
  arXiv:1002.4278 [hep-th];
\bibitem{iCosmo5} U.~H.~Danielsson,
  ``Entropic dark energy and sourced Friedmann equations,''
  arXiv:1003.0668 [hep-th];
\bibitem{iCosmo6} J.~W.~Lee,
  ``Zero Cosmological Constant and Nonzero Dark Energy from Holographic
  Principle,''
  arXiv:1003.1878 [hep-th];
\bibitem{iCosmo7} Y.~F.~Cai, J.~Liu and H.~Li,
  ``Entropic cosmology: a unified model of inflation and late-time
  acceleration,''
  arXiv:1003.4526 [astro-ph.CO];
\bibitem{iCosmo8} A.~Sheykhi,
  ``Entropic Corrections to Friedmann Equations,''
  arXiv:1004.0627 [gr-qc];
\bibitem{iCosmo9} R.~G.~Cai and S.~P.~Kim,
  ``First law of thermodynamics and Friedmann equations of
  Friedmann-Robertson-Walker universe,''
  JHEP {\bf 0502}, 050 (2005)
  [arXiv:hep-th/0501055];
\bibitem{iCosmo10} S.~W.~Wei, Y.~X.~Liu and Y.~Q.~Wang,
  ``Friedmann equation of FRW universe in deformed Horava-Lifshitz gravity from
  entropic force,''
  arXiv:1001.5238 [hep-th];
\bibitem{iCosmo11} M.~Li and Y.~Pang,
 ``A No-go Theorem Prohibiting Inflation in the Entropic Force Scenario,''
  arXiv:1004.0877 [hep-th];
\bibitem{iCosmo12} S.~W.~Wei, Y.~X.~Liu and Y.~Q.~Wang,
  ``Friedmann equation of FRW universe in deformed Horava-Lifshitz gravity from
  entropic force,''
  arXiv:1001.5238 [hep-th].
\bibitem{iNewton}
C.~Gao,
  ``Modified Entropic Force,''
  arXiv:1001.4585 [hep-th];
\bibitem{iNewton1} S.~Ghosh,
  ``Planck Scale Effect in the Entropic Force Law,''
  arXiv:1003.0285 [hep-th];
\bibitem{iNewton2} I.~V.~Vancea and M.~A.~Santos,
  ``Entropic Force Law, Emergent Gravity and the Uncertainty Principle,''
  arXiv:1002.2454 [hep-th];
\bibitem{iNewton3} L.~Modesto and A.~Randono,
  ``Entropic corrections to Newton's law,''
  arXiv:1003.1998 [hep-th];
\bibitem{iNewton4} M.~R.~Setare and D.~Momeni,
  ``Revisiting the Entropic corrections to Newton's law,''
  arXiv:1004.2794 [physics.gen-ph];
\bibitem{iNewton5} Y.~Zhao,
  ``Entropic force and its fluctuation in Euclidian quantum gravity,''
  arXiv:1002.4039 [hep-th].
\bibitem{iRindler}
H.~Culetu,
  ``Boundary stress tensors for spherically symmetric conformal Rindler
  observers,''
  arXiv:1001.4740 [hep-th];
\bibitem{iRindler1}
 H.~Culetu,
  ``Comments on 'On the Origin of Gravity and the Laws of Newton', by Erik
  Verlinde,''
  arXiv:1002.3876 [hep-th];
\bibitem{iRindler2}
J.~W.~Lee,
  ``On the Origin of Entropic Gravity and Inertia,''
  arXiv:1003.4464 [hep-th];
\bibitem{iRindler3} I.~V.~Vancea and M.~A.~Santos,
  ``Entropic Force Law, Emergent Gravity and the Uncertainty Principle,''
  arXiv:1002.2454 [hep-th].
\bibitem{specific}
 R.~G.~Cai, L.~M.~Cao and N.~Ohta,
  ``Notes on Entropy Force in General Spherically Symmetric Spacetimes,''
  Phys.\ Rev.\  D {\bf 81}, 084012 (2010)
  [arXiv:1002.1136 [hep-th]];
\bibitem{specific1} R.~A.~Konoplya,
  ``Entropic force, holography and thermodynamics for static space-times,''
  arXiv:1002.2818 [hep-th];
\bibitem{specific2}  P.~Mahato,
``Axial Current, Killing Vector and Newtonian Gravity,''
  arXiv:1004.1818 [gr-qc].
\bibitem{iLQG}
L.~Smolin,
  ``Newtonian gravity in loop quantum gravity,''
  arXiv:1001.3668 [gr-qc];
\bibitem{iLQG1}
F.~Caravelli and L.~Modesto,
  ``Holographic actions from black hole entropy,''
  arXiv:1001.4364 [gr-qc];
\bibitem{iLQG2} Yu-Xiao Liu, Yong-Qiang Wang and Shao-Wen Wei,
``Temperature and Energy of 4-dimensional Black Holes from
Entropic Force'', arXiv:1002.1062[hep-th].
\bibitem{Maekela}
J.~Makela,
  ``Notes Concerning 'On the Origin of Gravity and the Laws of Newton' by E.
  Verlinde (arXiv:1001.0785),''
  arXiv:1001.3808 [gr-qc].
\bibitem{otherPadmanabhan}
T.~Padmanabhan,
  ``Gravitational entropy of static spacetimes and microscopic density of
  states,''
  Class.\ Quant.\ Grav.\  {\bf 21}, 4485 (2004)
  [arXiv:gr-qc/0308070].
\bibitem{Strominger-Vafa}
  A.~Strominger and C.~Vafa,
  ``Microscopic Origin of the Bekenstein-Hawking Entropy,''
  Phys.\ Lett.\  B {\bf 379}, 99 (1996)
  [arXiv:hep-th/9601029].
\bibitem{Cardy}
J.~L.~Cardy,
  ``Operator Content Of Two-Dimensional Conformally Invariant Theories,''
  Nucl.\ Phys.\ B {\bf 270}, 186 (1986);
\bibitem{Cardy1}
  H.~W.~J.~Bloete, J.~L.~Cardy and M.~P.~Nightingale,
  ``Conformal Invariance, The Central Charge, And Universal Finite Size
  Amplitudes At Criticality,''
  Phys.\ Rev.\ Lett.\  {\bf 56}, 742 (1986).
\bibitem{Carlip}
  S.~Carlip,
  ``Black Hole Thermodynamics and Statistical Mechanics,''
  Lect.\ Notes Phys.\  {\bf 769}, 89 (2009)
  [arXiv:0807.4520 [gr-qc]];
\bibitem{Carlip1}
S.~Carlip,
  ``Symmetries, Horizons, and Black Hole Entropy,''
  Gen.\ Rel.\ Grav.\  {\bf 39}, 1519 (2007)
  [Int.\ J.\ Mod.\ Phys.\  D {\bf 17}, 659 (2008)]
  [arXiv:0705.3024 [gr-qc]];
\bibitem{Carlip2}
S.~Carlip,
  ``Conformal field theory, (2+1)-dimensional gravity, and the BTZ black
  hole,''
  Class.\ Quant.\ Grav.\  {\bf 22}, R85 (2005)
  [arXiv:gr-qc/0503022];
\bibitem{Carlip3}
S.~Carlip,
  ``Black Hole Entropy and the Problem of Universality,''
  arXiv:0807.4192 [gr-qc].
\bibitem{Strominger}
A.~Strominger,
  ``Five Problems in Quantum Gravity,''
  Nucl.\ Phys.\ Proc.\ Suppl.\  {\bf 192-193}, 119 (2009)
  [arXiv:0906.1313 [hep-th]].
\bibitem{Morozov-kin}
  A.~Morozov,
  ``Black Hole Motion in Entropic Reformulation of General Relativity,''
  arXiv:1003.4276 [hep-th].
\bibitem{other} Y.~Tian and X.~Wu,
  ``Thermodynamics of Black Holes from Equipartition of Energy and
  Holography,''
  arXiv:1002.1275 [hep-th].


\end{thebibliography}
\end{document}